\documentclass[12pt]{iopart}

\usepackage{graphicx}
\usepackage{dcolumn}
\usepackage{bm}
\usepackage{dsfont}
\usepackage[hypertex]{hyperref}

\newcommand{\R}{\mathds R}

\begin{document}

\title[Dimensional dependence of naked singularity formation]{Dimensional dependence of naked singularity formation in spherical gravitational collapse}

\author{Roberto Giamb\`o}
\ead{roberto.giambo@unicam.it}
\address{
Dipartimento di Matematica e Informatica\\
Universit\`a di Camerino\\ 62032 Camerino (ITALY)}%
\author{Sara Quintavalle}
\ead{sara.quintavalle@unicam.it}
\address{International School of Advanced Studies\\
Universit\`a di Camerino\\ 62032 Camerino (ITALY)}

\begin{abstract}
The complete spectrum of the endstates - naked singularities, or blackholes - of gravitational collapse is analyzed for a wide class of $N$-dimensional spacetimes in spherical symmetry, which includes and generalizes the dust solutions and the case of vanishing radial stresses.
The final fate of the collapse is shown to be fully determined by the local behavior of a single scalar function and by the dimension $N$ of the spacetime. In particular, the ``critical'' behavior of the $N=4$ spacetimes, where a sort of phase transition from black hole to naked singularity can occur, is still present if $N=5$ but does not occur if $N > 5$, independently from the initial data of the collapse.
Physically, the results turn out to be related to the kinematical properties of the considered solutions.
\end{abstract}


\maketitle

\section{Introduction}

Understanding singularities has always been one of the most intriguing issues in General Relativity since its beginning. The mathematical prediction that gravitational collapse may lead to singularity formation hugely increased the attention over the study of last stage of heavy stars' life. Problems related to strong density regions of spacetimes are in need of an ultimate answer already and, over all, a satisfactory formulation of Penrose's Cosmic Censorship Conjecture \cite{pe}: the causal character, and the endstate, of singularities arising from a dynamical process such as an indefinite collapse, is still one of the favorite test--bed for relativity.
A great amount of work in this direction has been done in the case of spherically symmetric 4--dimensional spacetimes: a number of collapsing models have been analytically studied where, under suitable assumptions, the arising singularity is not completely hidden behind a horizon, also when the latter forms. For instance the pioneering work of Christodoulou \cite{ch1} showed that it suffices removing homogeneity assumption from the paradigm of gravitational collapse leading to black hole -- i.e. Oppenheimer--Snyder solution. These cases of \emph{naked} singularities have been intensively explored, in particular Tolman--Bondi--Lemaitre dust clouds (see \cite{jbook} and references therein), and vanishing radial stress models \cite{gm,ha}.

Recently, a class of new solutions have been found out \cite{ggmp}, including the above as particular cases, where naked singularities generically appears as an outcome of collapse. Physically, they describe the gravitational
collapse of a class of anisotropic elastic materials, and are characterized by a particular choice of the equation of state that, in a certain coordinate system, allows to reduce Einstein Field Equations to a quadrature. In this paper, we find a natural extension of this class of solutions to the case of general $N$--dimensional gravitation theory. The importance of higher dimensional models goes up e.g. to
Kaluza-Klein theories, superstring theory, and brane--world models -- see in particular \cite{gre,rs}, where a description of the world with more than four non compact dimensions is proposed.

In this perspective, the present study is motivated by a number of earlier and more recent works on spherically symmetric higher dimensional spacetimes: \cite{Bl,LM} extend earlier well--known results and properties of the four dimensional scalar field collapse;  Vaidya--adS four dimensional solution is generalized in \cite{Ma,NM} to higher dimensions adding extra gravity terms to the action functional. Far from being exhaustive, more references on the subject of higher dimensional
collapse are \cite{barrow,gho,gho2,joshi2007,mae}.
In particular, the class of solutions that we find extend again vanishing radial stress models as dust \cite{chakra,pat}. We will find the complete spectrum of
endstates, analyzing if and how it is modified by the dimension
of the spacetime $N$. In particular,  naked singularities will be proved to survive in any larger dimension, despite earlier results contained in \cite{joshi1,joshi2} -- see discussion at the end.

It is worth noticing that some criticism arose to singularities occurring in  astrophysical sources modeled with continuous media in the past, due to the fact that one can construct situations in which Newtonian systems made out of continua develop singularities. As a consequence,
singularities in these models cannot be considered as an exclusive product of General Relativity. It is difficult, however, to assess to which extent this phenomenon denies validity to continuous models, although a simple remark once made by H. Seifert \cite{seif} may be of help: on taking this point of view, one could discard the big-bang of the standard model as being an artifact of Newtonian gravity, since Friedmann equation holds - formally unchanged - also for the Newtonian cosmological models.

The paper is organized as follows: section \ref{sec:solution} is devoted to derive and the class of exact solutions, and to illustrate briefly some particular cases. Physical reasonability conditions will also be imposed to the solution, together with conditions that will ensure formation of singularities, whose endstate will be analyzed in section \ref{sec:endstate}. In section \ref{sec:matching} we will show how to complete the  model, matching the
solution to a suitable exterior spacetime. In the final section we discuss the results found, relating them to kinematical properties of the spacetime.

\section{The solution in area--radius coordinates}\label{sec:solution}

The general spherically symmetric line element in comoving coordinates $(t,r,\theta^i)$, $i=1,\ldots,N-2$, is given by
\begin{equation}\label{eq:ds-com}
\mathrm ds^2=-e^{2\nu(t,r)}\,\mathrm dt^2 + {\eta(t,r)}^{-1}\,\mathrm dr^2 + R(t,r)^2\,\mathrm d\Omega^2_{N-2},
\end{equation}
where $\mathrm d\Omega^2_{N-2}\equiv\sum_{i=1}^{N-2}\Bigl(\prod_{j=1}^{i-1}\sin^2\theta^j\Bigr)(\mathrm d\theta^i)^2$.
The source of the gravitational field will be given by an elastic material in isothermal conditions. Generalizing the $N=4$ case , the property of the source are encoded in a state function depending on the \emph{space--space} part of the metric, that is -- using spherical symmetry assumption --  $w=w(r,R,\eta)$ \cite{km,gm}.
The stress energy tensor is given by
\begin{equation}\label{eq:T}
T=-\epsilon \,\mathrm dt\otimes\frac{\partial}{\partial t}+ p_r
\mathrm dr\otimes\frac{\partial}{\partial
r}+p_t\Bigl(\mathrm d\theta^i\otimes\frac{\partial}{\partial\theta^i}\Bigr),
\end{equation}
where, introducing the matter density
$$\rho=(N-2)(8\pi E)^{-1}\sqrt\eta\, R^{-2}$$
 ($E$ is an arbitrary function of $r$),  the internal energy $\epsilon$ and the stresses $p_r$ and $p_t$ are given in terms of the state function by
\begin{equation}\label{eq:3}
\epsilon=\rho\,w,\,\quad p_r=2\rho\eta\frac{\partial w}{\partial \eta}  ,\,\quad
p_t=-\frac{1}{N-2}\rho R\frac{\partial w}{\partial R}.
\end{equation}

Although the comoving coordinates usually yields the natural system to describe the physical evolution of the collapse, for our purposes, however, it will be convenient to introduce the \emph{area--radius} coordinate system $(r,R,\theta^i)$, first introduced by Ori \cite{ori} in the study of 4-dimensional charged dust, in such a way that \eref{eq:ds-com} becomes
\begin{equation}\label{eq:ori}
\mathrm ds^2=-A\,\mathrm dr^2 -2B\,\mathrm dR\,\mathrm dr -u^{-2}\,\mathrm dR^2 + R^2\,\mathrm d\Omega^2_{N-2},
\end{equation}
with $A,B,u$ unknown functions of $(r,R)$.
In this way the internal energy will depend only on one field variable, $\eta$, and on the two coordinates $r,R$. We introduce the function
$$\Delta=B^2-Au^{-2}=\eta^{-1}u^{-2},$$
so that Einstein field equations $G^r_r=8\pi T^r_r$, $G^R_r=8\pi T^R_r$ and $G^r_R=8\pi T^r_R$ can be expressed in terms of $A, \Delta$ and $u$ as follows:
\begin{eqnarray}
(1-\frac N2)[(N-3)(1-{A}/{\Delta})-R({A}/{\Delta})_{,R}]=8\pi R^2 p_r ,\label{eq:Grr}\\
(1-\frac N2)R^{-1}(A/\Delta),_r=-8\pi \sqrt{\Delta+A u^{-2}}u^{-2}(\epsilon+p_r), \label{eq:GRr}\\
\sqrt{u^2+A/\Delta}(\sqrt\Delta)_{,R}-(u^{-1})_{,r}=0. \label{eq:GrR}
\end{eqnarray}
Equation \eref{eq:Grr} can be integrated to give $A/\Delta$ in terms of $p_r(r,R,\eta)$.
Therefore, if one removes dependency on the comoving field variables, assuming that $p_r$ in \eref{eq:3} satisfies
$$\frac{\partial p_r(r,R,\eta)}{\partial\eta}=0,$$
or equivalently
\begin{equation}\label{eq:5}
w=h(r,R)+\ell(r,R)\eta^{-1/2}
\end{equation}
with $h,\ell$ arbitrary, then one obtains
\begin{equation}\label{eq:A}
A=\Delta(1-2\Psi\,R^{3-N})
\end{equation}
where, in view of \eref{eq:3} and \eref{eq:5}, $\Psi$ is the function
\begin{equation}\label{eq:6}
\Psi(r,R)=F(r)+\frac1{E(r)}\int_{R_0(r)}^R\ell(r,\sigma)\,\mathrm d\sigma,
\end{equation}
with $F(r)$ arbitrary function of $r$, and $R_0(r)$ describing $r$ at initial (comoving) time, that will be chosen equal to $r$ hereafter. The function $\Psi$ \eref{eq:6} is  Misner--Sharp mass of the system, defined by the relation $1-2\Psi R^{3-N}=g(\nabla R,\nabla R)$.
Now, inroducing
\begin{equation}\label{eq:Y}
Y(r,R)=E(r)\,{\Psi,_r(r,R)}\,{h(r,R)}^{-1},
\end{equation}
the field equations \eref{eq:GRr}--\eref{eq:GrR}, in view of \eref{eq:3}, \eref{eq:5} and \eref{eq:6}, simply become respectively
\begin{equation}
u^2=2\Psi R^{3-N}-1+Y^2,\label{eq:u}
\end{equation}
and $(\sqrt\Delta)_{,R}+Y^{-1}(u^{-1})_{,r}=0$, that can be integrated, using the initial condition, to give
\begin{equation}\label{eq:sqrtDelta}
\sqrt\Delta(r,R)=\int_r^R\frac{u_{,r}(r,\sigma)}{Y(r,\sigma)u(r,\sigma)^2}\,\mathrm d\sigma+\frac1{Y(r,r)u(r,r)}.
\end{equation}
Then, we conclude that the class of exact solutions found expresses all the metric unknown functions in \eref{eq:ori} in terms of two arbitrary functions $(\Psi, Y)$  of $r$ and $R$.

We stress the fact that
the constitutive function $w(r,R,\eta)$ as equation of state, introduced at the beginning of this section, \emph{uniquely} and \emph{completely} carries on the physical properties of the matter, regardless of possible anisotropies. Isotropy of the matter is characterized when $w$ can be written as a function of the matter density $\rho$ only. In this case, $p_r=p_t$ and both can be seen as a function of the energy density $\epsilon$ only, as one can easily calculate from \eref{eq:3}. When $w$ fails to be a function of $\rho$ only, anisotropy comes into play, but it is not needed any other relation to close the system, because of equations \eref{eq:3}. Another way to see this is to observe that equations (3) identically imply the conservation law arising from one of the Bianchi identities written in comoving coordinates, and again this is of course an outcome of having assumed that the source is an elastic continuum in isothermal conditions.  Of course, the requirement given by \eref{eq:5} is exactly the state function characterizing the class of function considered, and the fact that the arbitrary functions can be viewed in terms of the kinematical properties of the continuum is a very well known consequence of the structure of the field equations within the assumed symmetries and holds for all the models of this kind.

\subsection{Examples}
The components of the stress energy tensor are generically nonzero, as readily calculated from \eref{eq:3}--\eref{eq:6}, and are given by
\begin{eqnarray}
\epsilon=\frac{N-2}{8\pi
R^{N-2}}\Bigl[\frac{\sqrt{\eta}\Psi_{,r}}{Y}+\Psi_{,R}\Bigr],\label{eq:epsilon}\\
p_r=-\frac{(N-2)\Psi_{,R}}{8\pi R^{N-2}},\label{eq:pr}\\
p_t=-\frac{\sqrt{\eta}}{8\pi
R^{N-3}}(\frac{\Psi_{,rR}}{Y}-\frac{\Psi_{,r}}{Y^2}\frac{\partial
Y}{\partial R}+\frac{\Psi_{,RR}}{\sqrt{\eta}}).\label{eq:pt}
\end{eqnarray}
From these expressions we can recognize some particular cases:
\begin{enumerate}
\item dust spacetimes \cite{joshi1}, occurring when both $\Psi$ and $Y$ are functions of $r$ only;
\item vanishing radial stress solutions \cite{joshi2}, that occur when $\Psi$ is a function of $r$ only, but $Y$ may also depend on $R$,
\item acceleration free solutions, when $Y$ is a function of $r$ only but $\Psi$ may also depend on $R$ (note that the norm of the acceleration is simply given by $Y_{,R}$)
\end{enumerate}

\subsection{Energy condition and shell focussing singularity occurrence}
On the above class of solutions, some conditions will be imposed, as requirements on $\Psi$ and $Y$. Since we want to obtain global gravitational collapsing models, we will consider the interior metric \eref{eq:ori} as defined on a right neighborhood $[0,r_b]$ of $r=0$, for some $r_b>0$, and match the above solutions at $r=r_b$ with some exterior spacetime to be defined later (see section \ref{sec:matching}).
For this reason, in the following we will consider $\Psi$ and $Y$ as defined on the set $\{(r,R)\,:\,r\in[0,r_b],\,R\in[0,r]\}$.

As a physical reasonability condition, WEC on the metric \eref{eq:ori} will be required, but in view of \eref{eq:epsilon}--\eref{eq:pt}, it suffices that
\begin{eqnarray*}
\Psi,_r\ge 0,\qquad\Psi,_R\ge 0\\
(N-2) \Psi,_r Y^{-1}\ge R (\Psi,_r Y^{-1}),_R,\qquad (N-2)\Psi,_R\ge R\Psi_{RR}
\end{eqnarray*}
Moreover, we impose the condition of decreasing initial energy, i.e. $\epsilon_0(r):=\epsilon(r,r)$ must be a decreasing function of $r$:
$$
\Psi_{,rr}(r,r)+\Psi_{,RR}(r,r)+2\Psi_{,rR}(r,r)
\le\frac{(N-2)}{r}[\Psi_{,r}(r,r)+\Psi_{,R}(r,r)].
$$
The functions $\Psi$ and $Y$ must be chosen in such a way that the spacetime is regular at initial (comoving) time, and a (shell focussing) singularity forms, for each shell $r\in[0,r_b]$,  in a finite amount of time. Therefore, first of all shell crossing singularity formation must be avoided, and to this aim it must be required that $\sqrt\Delta>0$ when $R\ge 0$. By inspection of \eref{eq:sqrtDelta},  sufficient conditions for this to happen are given by
$$
\sqrt\Delta(r,0)>0,\qquad u_{,r}(r,R)>0,\qquad\forall r\in]0,r_b],\,R\in]0,r].
$$

Moreover, it must be observed that the use of the $(r,R,\theta^i)$ coordinate system has the obvious advantage to parameterize the singularity with the straight line $R=0$, but the drawback that both the regular and the singular centre are mapped into the point $r=R=0$, and then it does not make a distinction between them, unless one does not consider the inverse function $t=t(r,R)$. The function  $\dot R$, the derivative of $R$ w.r.t comoving time, satisfies the identity $u=-\dot R e^{-\nu}$, that can be formally integrated to give
$t(r,R)=\int_R^r e^{-\nu}u(r,\sigma)\,\mathrm d\sigma$. Although the integrand yet contains an unknown function in the comoving coordinates, a key remark at this stage is to observe that $e^{-\nu}$ is bounded in a neighborhood of the centre, which allows to express the above conditions in terms of $\Psi$ and $Y$:
it suffices that the function $R^{N-3}u^2$ is Taylor expandable at the centre $(r=R=0)$, with expression given by
\begin{equation}\label{eq:10}
R^{N-3}u^2=\sum_{i+j=N-1}h_{ij}r^iR^j+\sum_{i+j=N-1+p}h_{ij}r^iR^j+\ldots
\end{equation}
In particular, for the centre to become singular in a finite time, it must be required that $(h_{N-1,0},h_{N-2,1},\ldots,h_{1,N-2})\ne 0$. Hereafter, we will suppose, as already done in \cite{ggmp},
\[
\alpha:=h_{N-1,0}\ne 0.
 \]
Although this is a generic assumption, the results we are going to state can also be extended to the degenerate case $\alpha=0$, as done in \cite{W} for $N=4$.

\section{Naked singularity vs. black hole formation}\label{sec:endstate}
The endstate of the singularity for these models will be studied.  First, let us observe that the central singularity is the only one that can be naked. Indeed, under the above assumptions, the apparent horizon $R_h(r)$ is such that $R_h(r)= \alpha^{\frac{1}{N-3}}
r^{\frac{N-1}{N-3}}+o(r^{\frac{N-1}{N-3}})$, and moreover, if $t_h(r)$ and $t_s(r)$ are the comoving times when the shell labeled $r$ becomes trapped and singular, respectively, then $\lim_{r\to 0^+}t_h(r)-t_s(r)=0$.

To analyze the endstate of the central singularity we will study existence of null radial geodesics $R_g(r)$ emanating from the (singular) centre, such that $R_g(r)>R_h(r)$ in a right neighborhood of $r=0$. To do that, we will use a remarkable property of $R_h(r)$, to be a \emph{supersolution} of null radial geodesic equation
\begin{equation}\label{eq:geo}
\frac{\mathrm dR}{\mathrm dr}=u\sqrt{\Delta}(Y - u).
\end{equation}
Therefore, to have existence of such a $R_g(r)$ as above, we will actually look for \emph{subsolutions} of \eref{eq:geo} of the form $R_x(r)=xr^{\frac{N-1}{N-3}}$, with $x>\alpha^{\frac{1}{N-3}}$, that therefore emanate from the singular centre - so that $R_g$ also will. Incidentally, this also explains why it suffices to look for \emph{radial} curves: indeed, the projection of a nonradial geodesic on the $(r,R)$ plane would be a supersolution of \eref{eq:geo}, so if the singularity is nonradially naked, is also radially naked.

As it happens for the $N=4$ case, the endstate of the singularity is related to Taylor expansion of the function
\begin{equation}\label{eq:12}
\sqrt\Delta(r,0)=\xi r^{n-1}+o(r^{n-1}),
\end{equation}
but  also the dimension $N$ of the spacetime will play now a crucial role. Indeed, the condition for the existence of $R_x$ as above is equivalent to the existence of $x>\alpha^{\frac{1}{N-3}}$ satisfying
\begin{equation}\label{eq:root}
\frac{N-1}{N-3}xr^{\frac{2}{N-3}} <
(1-\sqrt{\frac{\alpha}{x^{N-3}}})(\sqrt{\frac{\alpha}{x^{N-3}}}\xi
r^{n-1}+xr^{\frac{N(N-3)}{2}}).
\end{equation}
The above inequality gives the complete spectrum of the endstates since it provides a necessary and sufficient condition for the singularity to be naked.
Indeed,
if $N=4$ one recovers the well known results of \cite{ggmp}
that the inequality holds -- and hence the singularity is naked -- if $n=1,2$, and if $n=3$ a critical case happens when the endstate is related to the value of $\xi$  in \eref{eq:12}, since it must be $2\xi>(26+5\sqrt 3)\alpha$ for the singularity to be naked.
\emph{In larger dimensions, the singularity is naked if $n=1$, $\forall N$, and if $n=2,N=5$, provided $2\xi>27\sqrt\alpha$. In all other cases a black hole forms.} Then we observe that the critical behavior, when a phase transition from black hole to naked singularity occurs, depending on the value of $\xi$, is a feature of dimensions $N=4$, and $N=5$, and is forbidden at larger dimensions. As one can see, the contribution of the dimension $N$, when it is larger than four, basically enters in the behavior of the apparent horizon, that behaves like $r^{1+2/(N-3)}$, which is no more an integer power of $r$ as $N\ge 6$, and it is always leading upon the ``kinematical'' contribution of $N$ -- i.e. the last term in \eref{eq:root}. Since the contribution of $\sqrt\Delta(r,0)$ is always an integer power of $r$ -- see below -- this fact results in the lack of critical case when $N\ge 6$.

\begin{figure}
\begin{center}
\includegraphics[width=8cm,height=6cm]{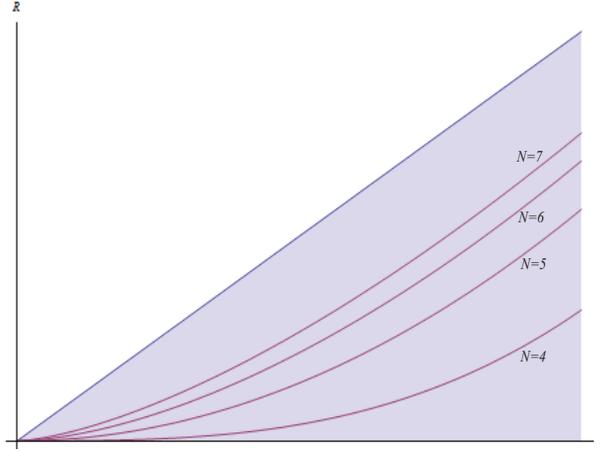}
\end{center}
\vskip-0.5cm \caption{Asymptotic behavior of the apparent horizon with respect to the dimension near the centre. In the $(r,R)$ plane, the shaded region represents the evolution of the solution. Fixed the value for $\alpha$, the higher $N$ is, the bigger is the trapped region lying between the $r$--axes and the horizon.} \label{fig2}
\end{figure}

\section{Exterior spacetime and matching conditions}\label{sec:matching}

In this section we will see how to complete the model, matching the interior solution studied so far with an exterior spacetime, and requiring that Israel--Darmois junction conditions hold along the matching hypersurface $\Sigma=\{r=r_b\}.$ From \eref{eq:pr} we observe that radial pressure $p_r$ does not vanish in general along $\Sigma$, so we cannot expect to match the solution with a Schwarzschild exterior. In this case a natural choice for the
exterior metric can be given by generalized Vaidya solutions \cite{joshi2007,ww}, that for generic $N$ read
$$
\mathrm ds_{\mathrm{ext}}^2=-\left(1-\frac{2M(V,R)}{R^{N-3}}\right)\,\mathrm dV^2-2\,\mathrm
dR\,\mathrm dV + R^2\,\mathrm d\Omega_{N-2}^2,
$$
and Israel--Darmois junction conditions simply become requirements on the mass function $M(V,R)$ on the junction hypersurface. To find the conditions, it is convenient to work with the general interior metric written in comoving coordinates \eref{eq:ds-com}. Parameterized $\Sigma$ with coordinates $(\tau,\theta^i)\hookrightarrow(\tau,r_b,\theta^i)$, The first and second fundamental forms of $\Sigma$ w.r.t. this metric read
\begin{eqnarray}
\mathds I_{\mathrm{int}}^\Sigma=-e^{2\nu}\,\mathrm d\tau^2+R^2\mathrm d\Omega^2_{N-2},\label{eq:Iint}\\
\mathds{II}_{\mathrm{int}}^\Sigma=-\eta^{1/2}\left(e^{2\nu}\nu'\mathrm d\tau^2-R\,R'\mathrm d\Omega^2_{N-2}\right),\label{eq:IIint}
\end{eqnarray}
where a dash and a dot denote derivatives w.r.t. $r$ and $t$ respectively, and all functions are intended evaluated in $(\tau,r_b)$.
Injection of $\Sigma$ into the exterior spacetime reads in coordinates as $(V(\tau),Y(\tau),\theta^i)$, where $V(\tau),Y(\tau)$ must be determined. The first fundamental form of $\Sigma$ takes the form
\begin{equation}\label{eq:Iext}
\mathds I_{\mathrm{ext}}^\Sigma=-\left[\left(1-\frac{2M(V(\tau),Y(\tau))}{Y(\tau)^{N-3}}\right)\dot V(\tau)^2+2\dot V(\tau)\dot Y(\tau)\right]\mathrm d\tau^2+Y(\tau)^2\mathrm d\Omega^2_{N-2},
\end{equation}
where, with a slight abuse of notation, we denote by a dot the derivative w.r.t. $\tau$.
Comparing \eref{eq:Iint} with \eref{eq:Iext} gives
\begin{eqnarray}
Y(\tau)=R(\tau,r_b),\label{eq:j1}\\
\left(1-\frac{2M}{Y^{N-3}}\right)\dot V^2+2\dot V\dot Y=e^{2\nu},\label{eq:j2}.
\end{eqnarray}
Using these relations we can express the second fundamental form of $\Sigma$ w.r.t. the exterior metric as
\begin{eqnarray}
\fl \mathds{II}_{\mathrm{ext}}^\Sigma=-e^{\nu}\left\{\left[-\dot Y\left(\ddot V-\frac12\chi_{,Y}\dot V^2\right)+\dot V\left(\ddot Y+\chi_{,Y}\dot Y\dot V+\frac12\chi\,\chi_{,Y}\dot V^2+\frac12\chi_{,V}\dot V^2\right)\right]\right.\mathrm d\tau^2
\nonumber\\
\left.-Y(\dot Y+\chi\dot V)\mathrm d\Omega^2_{N-2}\right\},\label{eq:IIext}
\end{eqnarray}
where  $\chi=1-2M Y^{3-N}$. Comparing angular terms in \eref{eq:IIint} and \eref{eq:IIext} and using \eref{eq:j2} gives
\begin{equation}\label{eq:jmass}
\chi=R'^2\eta-(\dot R e^{-\nu})^2,
\end{equation}
that is continuity of Misner--Sharp mass across $\Sigma$. Therefore we find the differential equation for $V(\tau)$:
\begin{equation}\label{eq:j3}
\dot V(\tau)=\frac{e^\nu}{R'\eta+\dot R e^{-\nu}}.
\end{equation}
At this stage, it remains to compare $\mathrm d\tau^2$ terms in the second fundamental forms. But, with some algebra, the above relations together with field equation $\dot R'=\dot\lambda R'+\nu'\dot R$ simply reduce the condition to
\begin{equation}\label{eq:j4}
{M_{,V}}(V(\tau),Y(\tau))=0.
\end{equation}
Therefore, we conclude that generalized Vaidya solutions can always be matched to a spherically symmetric interior metric \eref{eq:ds-com} along $\Sigma$, provided that conditions \eref{eq:j1} and \eref{eq:j3} hold, and the mass function $M(V,Y)$ satisfies \eref{eq:jmass} and $\eref{eq:j4}$ on $\Sigma$.

The above fact can easily be translated in area--radius formalism: it suffices to parameterize $\Sigma$ with coordinates $(\sigma,\theta^i)\hookrightarrow(r_b,\sigma,\theta^i)$. In this case the injection of $\Sigma$ in the exterior spacetime reads $(V(\sigma),\sigma,\theta^i)$, where $V(\sigma)$, in view of \eref{eq:j1}--\eref{eq:j4}, becomes
\begin{equation}\label{eq:j1-ar}
\frac{\mathrm dV}{\mathrm d\sigma}=\frac1{u(r_b,\sigma)(u(r_b,\sigma)-Y(r_b,\sigma))},
\end{equation}
and the mass function satisfies
\begin{equation}\label{eq:jmass-ar}
M(V(\sigma),\sigma)=\Psi(r_b,\sigma),\qquad M_{,Y}(V(\sigma),\sigma)=\Psi_{,R}(r_b,\sigma).
\end{equation}
It can be observed that an interesting subclass of the above exterior metric is given by the anisotropic generalizations of deSitter spacetime \cite{aniso}, which is obtained taking  $M=M(Y)$. Obviously, in this case condition \eref{eq:j4} is trivially satisfied, and \eref{eq:jmass-ar} simply reduces to require continuity of the mass across the junction hypersurface (see also \cite{hsf}).

\section{Discussion and conclusions}
There have been previous works trying to explain the endstate in terms
of the kinematical properties of the spacetime, in particular the shear at initial time \cite{joshi1,joshi2}. In the following we are going to address this point, relating the indices $n$,$\xi$ coming from \eref{eq:12} to \emph{all} kinematical properties (see also \cite{chakra,mena}). The function $\sqrt\Delta(r,0)$ can be split in the sum of $I_1(r)+I_2(r)$, where
\begin{eqnarray*}
&&I_1(r):=\frac{1}{Y(r,r)}\frac{\partial}{\partial
r}\int_0^r\,\frac1{u(r,\sigma)}\,\mathrm d\sigma ,\\
&&I_2(r):=\int_0^r\left(\frac{1}{Y(r,\sigma)}-\frac{1}{Y(r,r)}\right){\left(\frac1{u(r,\sigma)}\right)}_{,r}\,\mathrm d\sigma.
\end{eqnarray*}
The behavior of these quantities near the centre can be studied, to find that $I_1(r)=p a r^{p-1}+o(r^{p-1})$, where $p$ is given in \eref{eq:10}, and $a\in\R$ depends on the coefficients $h_{ij}$ of order $N-1$ and $N-1+p$.

Introduced the polynomials $P_k(\tau)=\sum_{j=0}^k h_{k-j,j}\tau^j$, the value of $a$ is given by
$$
a=-\int_0^1\frac{P_{N-1+p}(\tau)\tau^{1/(N-3)}}{2 P_{N-1}(\tau)^{3/2}}\,\mathrm d\tau.
$$
On the other side, $I_2(r)=b r^{q}+o(r^q)$, where $b\in\R$ and $q$ is the order of the first nonvanishing term of $Y,_R(r,R)$ expansion at the centre. Then, $n$ in \eref{eq:12} is given by the smallest between $p$ and $q+1$. Now, the shear of the solution can be controlled by the  scalar $$\sigma=\frac12\sigma^{\mu\nu}\sigma_{\mu\nu}=-u\sqrt{(N/2-1)/(N-1)}(\log(R(u\sqrt\Delta)^{-1})),_{R}$$
and on the initial slice $R=r$ behaves like $p\sigma_0 r^p+o(r^p)$, where $\sigma_0=\sqrt{(N/2-1)/(N-1)}{(P_{N-1+p}(1))}/{(4 P_{N-1}(1))^{1/2}}$.
We deduce that the asymptotic behavior of the shear near the regular centre can be responsible for the quantity $I_1(r)$ only, and does not even control the value of the parameter $a$ in the critical case $(n=2,N=5)$ -- not to tell that one can conceive cases when $\sigma_0=0$.
On the other side, the norm of the acceleration is given simply by $Y,_R$,  and then it rules the quantity $I_2(r)$, but again the knowledge of the initial acceleration could not be enough to establish the value of $q$. We can conclude that \emph{the evolutions} of both acceleration and shear influence the endstate of the gravitational collapse, but none of them can be considered as a stand--alone responsible, as the function $\sqrt\Delta(r,0)$ is, together with the dimension $N$.

We observe that, if $N\ge 6$, the singularity is naked only when $n=1$.
This is not in contrast with \cite{joshi1,joshi2}, where dust and vanishing radial stress solutions are shown to produce a black hole when $N\ge 6$. Indeed, the special case considered in those paper are acceleration free, or more generally such that $\sqrt\Delta(r,0)$ behaves like $I_1(r)$ anyway, and so the endstate is related to the first nonvanishing power of $R^{N-3}u^2$, after the $(N-1)$--th order. In the cases produced in \cite{joshi1,joshi2} the expansion for both $\Psi(r)/r^{N-1}$ and $Y^2$ is assumed to contain only even order terms, which excludes the possibility $n=1$. Instead, in the case studied in the present paper a more general situation is considered, when $R^{N-3}u^2$ may contain both odd and even order terms, but only terms of order $N-1+2k$, $k\in\mathds N$, when restricted on the initial slice $R=r$. In other words, solutions may be produced, when $\Psi/r^{N-1}$ and $Y^2$ are even at initial time, but later they evolve to allow also for odd order terms. All in all, the conclusion stated for $N=4$ in \cite{ggmp} is confirmed at higher dimensions, that the formation of naked singularities or black holes weakly depends on the initial data, but is essentially a local phenomenon, depending on the  Taylor expansion of a kinematical invariant near the centre. The contribution of the dimension is basically related to the behavior of the apparent horizon, that forbids occurrence of critical cases when $N\ge 6$, and restricts, but still allows for naked singularity formation at any dimension.

\end{document}